\documentclass[conference]{IEEEtran}

\IEEEoverridecommandlockouts
\usepackage{cite}
\usepackage[shortcuts,acronym]{glossaries}
\usepackage{xcolor}
\usepackage{amssymb}
\usepackage{url}
\usepackage[]{algorithm2e}

\usepackage[pdftex]{graphicx}
 \graphicspath{{./}{./}}
\usepackage{subcaption}

\usepackage{amsmath}

\newcommand{\mb}[1]{\mathbf{#1}}

\makeglossaries % generates the acronym list

\makeatletter
\def\ps@IEEEtitlepagestyle{
  \def\@oddfoot{\mycopyrightnotice}
  \def\@evenfoot{}
}
\def\mycopyrightnotice{
  {\footnotesize
  \begin{minipage}{\textwidth}
  \centering
  \copyright~2019 IEEE. Personal use of this material is permitted. Permission from IEEE must be obtained for all other uses, in any current or future media, including reprinting/republishing this material for advertising or promotional purposes,creating new collective works, for resale or redistribution to servers or lists, or reuse of any copyrighted component of this work in other works.
  \end{minipage}
  }
}

\newacronym{eot}{EOT}{Extended Object Tracking}
\newacronym{gw}{GW}{Gaussian-Wasserstein}
\newacronym{mc}{MC}{Monte-Carlo}
\newacronym{mem-ekf*}{MEM-EKF*}{Multiplicative Error Model Extended Kalman Filter*}
\newacronym{mmospa}{MMOSPA}{Minimum Mean Optimal Subpattern Assignment}
\newacronym{mmse}{MMSE}{Minimum Mean Square Error}
\newacronym{rmse}{RMSE}{Root Mean Square Error}

\author{Kolja Thormann and Marcus Baum% <-this % stops a space
\thanks{Kolja Thormann and Marcus Baum are with the Institute of Computer Science, University of Goettingen, Germany,
{\tt\small \{kolja.thormann, marcus.baum\}@cs.uni-goettingen.de}}%
}

\begin{document}
\title{Optimal Fusion of Elliptic Extended Target Estimates based on the Wasserstein Distance}

\maketitle

% Abstract
\begin{abstract}
This paper considers the fusion of  multiple estimates of a spatially extended object, where the object extent is modeled as an ellipse  parameterized by the orientation and semi-axes lengths. 
For this purpose, we propose a novel systematic approach that employs a distance  measure for ellipses, i.e., the Gaussian Wasserstein distance, as a cost function. We derive an explicit approximate expression for the Minimum Mean  Gaussian Wasserstein distance  (MMGW) estimate. Based on the concept of a MMGW estimator, we develop efficient methods for the fusion of  extended target estimates.
The proposed fusion methods are evaluated in a simulated experiment and the benefits of the novel methods are discussed.
\end{abstract}


%Start of Document==============================================
\section{Introduction}\label{sec_intro}
With the improvement of sensor technology, the original assumption that an object yields at most one measurement per time step does not hold true anymore. Due to higher sensor resolution or closer targets, like in the automotive case, one object can cover multiple sensor cells, resulting in multiple detections per time step. Therefore, it must be assumed that the object is not a point, but possesses an extent from which measurements originate. The problem of simultaneously tracking an objects kinematic parameters and extent is called \ac{eot}~\cite{Mihaylova2014,Granstroem2017}. Especially with noisy measurements, this poses several challenges. One way to improve the target estimates is by the usage of multiple sensors for competitive fusion. However, to fully utilize the information from multiple estimates, they need to be fused accordingly.

There exist several approaches on how to model the extent, like rectangles~\cite{Nilsson2016}, ellipses~\cite{Koch2008, Feldmann2010, Yang2018_arXiv, Brosseit2017}, or arbitrary, star-convex shapes, modelled via Fourier coefficients~\cite{J_TAES_Baum_RHM}, Gaussian processes~\cite{wahlstrom2015extended,Hirscher2016}, or splines~\cite{Kaulbersch2018_Fusion}.

\begin{figure}[h]
\centering
\includegraphics[width=0.38\textwidth]{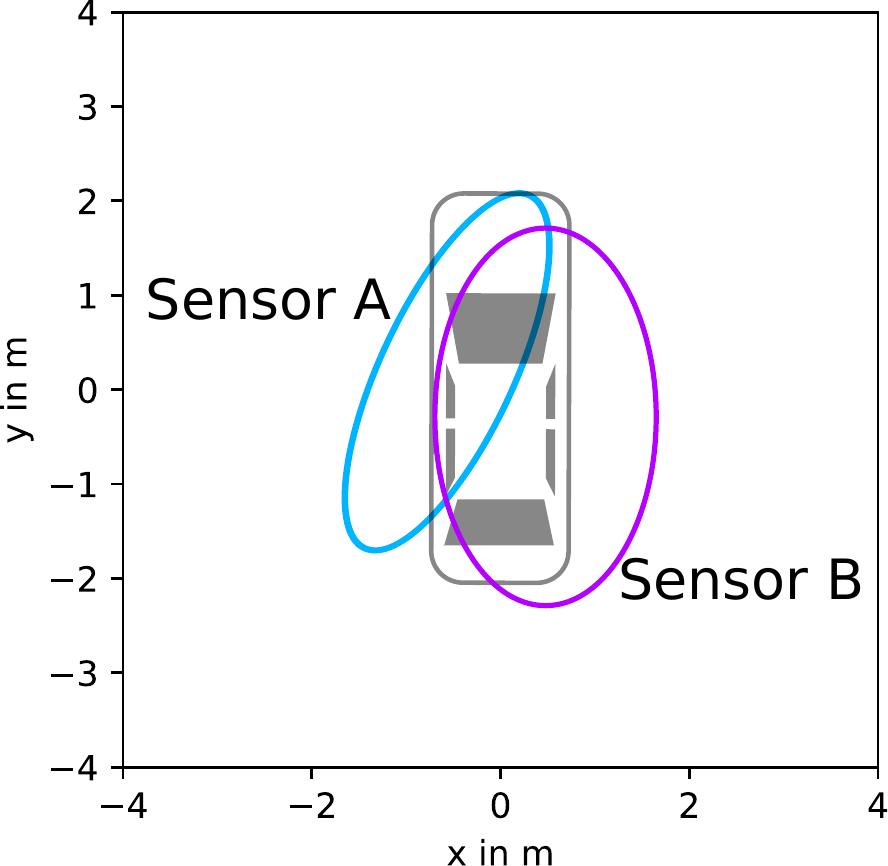}
\caption{A vehicle being tracked by two sensors A and B with the sensor estimates as a light blue and a purple ellipse.}
\label{fig_example}
\end{figure}

Ellipses can be utilized for tracking of ships~\cite{Granstroem2017} or pedestrians~\cite{Teich2017} or if the sensor noise is too high and the measurements are too sparse to get a precise estimate of the actual shape, as can be the case with radar. Ellipse trackers include the random matrix approach~\cite{Koch2008,Feldmann2010,Lan2016b,Vivone2015}, the \ac{mem-ekf*}~\cite{Yang2018_arXiv}, or the volcanormal model~\cite{Brosseit2017}. This paper will focus on ellipse extents and track-2-track fusion, which does not take the measurements from multiple sensor, but instead each sensor provides a track estimate. This way, raw sensor data stays locally and more compact information is sent to the centralized fusion unit~\cite{Durrant2008multisensor}. It must be noted that the state covariance must be provided by the individual sensor systems for track-2-track fusion to properly work~\cite{Michaelis2017heterogeneous}. We will discuss how to fuse ellipse estimates from multiple sensors appropriately (see Figure~\ref{fig_example}). The most important aspects to be considered here are the parametrization of the ellipses and the accuracies of the estimates.

This paper proposes a novel approach to fuse two elliptic extended target estimates  stemming from sensors with (possible) different accuracies. 
As an extended target estimate describes an uncertain geometric shape, the naive application of the standard Kalman filter formulas can yield to counter-intuitive results.
For this reason, we follow a novel systematic approach:
\begin{enumerate}
    \item A suitable distance measure that incorporates the geometric shape of the extended target, i.e., the \ac{gw} distance, is determined.
    \item A  cost function for extended target estimators, i.e., the Mean Gaussian Wasserstein (MGW) error, is defined. 
    \item  An \emph{explicit} approximation for the Minimum Mean Gaussian Wasserstein (MMGW) estimator, i.e., the optimal estimator with respect to the \ac{gw} distance,  is derived.
\item Practical algorithms for the fusion of two extended target estimates are developed based on the MMGW concept.
\end{enumerate}

The remainder of this paper is structured as follows. Section~\ref{sec_rel} presents related work in the area of sensor fusion. Next, the problem this paper intends to tackle is formulated in Section~\ref{sec_example}. Our approach is then explained in Section~\ref{sec_approach}. The actual fusion methods are presented in Section~\ref{sec_fusion}, followed by a heuristic alternative in Section~\ref{sec_bestOr}, and evaluated in Section~\ref{sec_eval}. Section~\ref{sec_concl} concludes the paper and outlines future work.

\section{Related Work}\label{sec_rel}
A widely-used approach to track elliptical objects is based on random matrices~\cite{Koch2008}. In \cite{granstrom2013spawning}, an intuitive method is proposed to combine two target estimates in a way that would create a fusion which was based on the joined measurement set of the individual sensors. There also exist approaches to combine measurements from multiple sensors for one target update~\cite{Vivone2017}, using the random matrix update by~\cite{Feldmann2010}, which is however not object-level fusion we are concerned with. Furthermore, these approaches do  not explicitly  incorporate covariance information for the estimated shape.

For star-convex shapes, \cite{Michaelis2017heterogeneous} propose a framework for multi-sensor tracking using Gaussian Processes. However, as the previous approaches they utilize the sensors' measurements directly. Another approach applying the Kullback-Leibler Divergence between two Random Finite Sets is proposed in~\cite{Frohle2019decentralized}.

An approach for extended object fusion of rectangular shapes is presented in~\cite{Nilsson2016}. They combine the corners of two rectangles using the Mahalanobis distance to incorporate different covariance matrices for the corner positions. \cite{Duraisamy2016track}~also propose a method to fuse shape estimates represented as segments with up to three points (point-, I-, or L-shape).

%% Das sind unsere Ideen für die Zukunft, nicht verraten!
%Formulating the shape fitting as an association between predetermined or sampled surface points, one could also consider the \ac{mmospa} estimator~\cite{Guerriero2010} as a possibility.

In \cite{MFI16_Yang}, several metrics suitable for elliptic \ac{eot} are discussed, including Intersection-over-Union~\cite{Granstroem2011}, the \ac{gw} distance~\cite{Givens1984}, and the Hausdorff distance, e.g.~\cite{Veltkamp2000}. It is pointed out that decoupling of state parameters is one option, as, e.g., calculating the Euclidean distance between two states could be difficult to interpret if the state consists of parameters in different units, like meters and radians. They compare \ac{gw} distance, Kullback-Leibler Divergence, and discretized OSPA distance in three scenarios, one where only the ellipse center differs, one where one ellipse is tilted and the semi-axes lengths differ, and one where only the orientation differs. They argue that for elliptic shapes, the \ac{gw} distance is most suitable for obtaining a single scalar score to rate the distance regarding all state parameters, as it can be solved in closed-form, provides intuitive results, and fulfills the properties of a metric.

This work is also inspired by the Minimum Mean OSPA estimators \cite{Guerriero2010,Crouse2011e,Crouse2013,J_SPL15_Baum} for multiple target tracking \cite{Vo2015}, where optimal estimates with respect to the OSPA distance \cite{Schuhmacher2008} are calculated.
In this context, the concept of a Wasserstein barycenter~\cite{Agueh2011,Bigot2012,Rabin2012,Carlier2014,Cuturi2013}
 is also related to this work.

\section{Motivating Example}\label{sec_example}
In this work, the spatial extent of a target is modelled as an ellipse. An ellipse can be specified by a five-dimensional vector $\mb{x} = [m_x,m_y,\alpha, l,w]^T$ that contains the  position $\mb{m}=[m_x,m_y]^T$, orientation $\alpha$, and semi-axis length $l$ and width $w$. This representation allows for explicitly maintaining different uncertainties for the shape parameters. This can become important, e.g., if we track a vehicle and only get measurements from the rear. In that scenario, we would have a high certainty about the ellipse's width, but a low certainty about its length.

\begin{figure}
\centering
\includegraphics[width=0.35\textwidth]{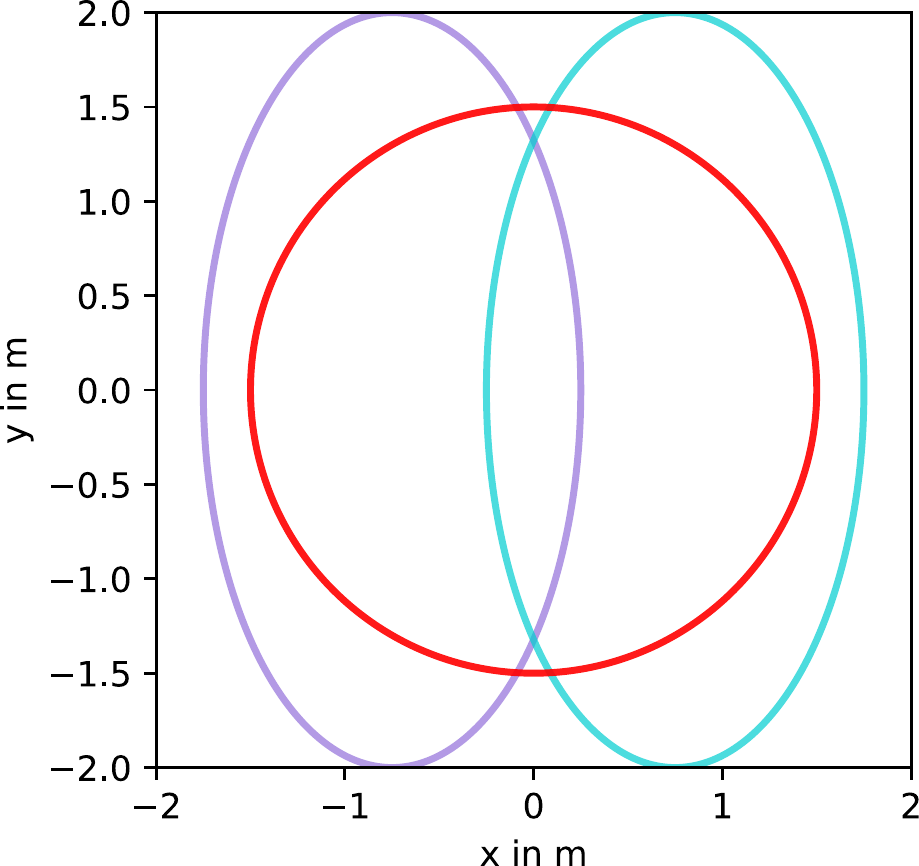}
\caption{Fusion of two ellipses using orientation, length, and width as representation. The left purple and right light blue ellipse posses the same shape, but the right is rotated by $\frac{\pi}{2}$ and has the axes switched. The middle red ellipse is the fusion of both.}
\label{fig_badexample}
\end{figure}

Now, assume two sensors A and B provide  the  two estimates $\mb{\hat{x}}_1$ and $\mb{\hat{x}}_2$ of an extended object.
In the most simple case of equal uncertainties, a linear fusion according to the Kalman filter would just give us the average
\begin{gather}\label{eqn:naivefusion}
\mb{\hat{z}}=\frac{1}{2}(\mb{\hat{x}}_1 + \mb{\hat{x}}_2)\enspace.
\end{gather}
This is the optimal fusion rule in a mean squared error sense based on the Euclidean distance for $x$.
However, this approach poses some serious weaknesses.  The problem is that multiple representations can be used to describe the same ellipse. Figure~\ref{fig_badexample} shows a scenario with equal shape estimates of an ellipse (they are apart for better visualization). The difference is that one ellipse has the values for $l$ and $w$ switched and the orientation is shifted by $\frac{\pi}{2}$. The resulting ellipse using  (\ref{eqn:naivefusion}) is  seriously counter-intuitive. This shows a different parametrization of the same ellipse actually leads to a different fusion result. 

To solve this issue, we must discuss the origin for  the above fusion formula (\ref{eqn:naivefusion}). Averaging the estimate in the Kalman filter comes from minimizing the mean squared error. Let $p(\mb{x})$ be a (posterior) probability density for $\mb{x}$. We then get the Minimum Mean Squared Error estimate (MMSE) $\mb{\hat{z}}$ as
\begin{gather}
\mb{\hat{z}}=\underset{\mb{z}}{\mathrm{argmin}}\int ||\mb{z} - \mb{x}||_2^2\cdot p(\mb{x})\,\text{d}\mb{x}=\mathbb{E}\{X\} \enspace .
\end{gather}
However, as shown in the example above, this is not suitable as the ellipse parameters describe a geometric object. For a systematic approach we have to replace the Euclidean distance with a distance metric on ellipses. We decided to utilize the \ac{gw} distance because, as mentioned in Section~\ref{sec_rel}, it can be argued to be  suitable to represent the distance between two ellipses with a single scalar value.

\section{Minimum Mean Gaussian Wasserstein Estimation}\label{sec_approach}
In order to calculate the Gaussian Wasserstein distance, 
we define the  shape matrix $\mb{X}$ of an ellipse as
\begin{align}
\mb{m}  &= [m_x,m_y]^{\text{T}}\\
\mb{X} &= \mb{R}_{\alpha}\cdot \mathrm{diag}(l^2, w^2)\cdot \mb{R}_{\alpha}^{\text{T}} \enspace
\end{align}
where   $\mb{R}_{\alpha}$ is a rotation matrix with angle $\alpha$.

The Minimum Mean Gaussian Wasserstein (MMGW) estimate for a probability distribution $p(\mb{x})$ is defined as  
\begin{gather}
\mb{\hat{z}}=\underset{\mb{z}}{\mathrm{argmin}}\int GW(\mb{n}, \mb{Z}; \mb{m}, \mb{X})\cdot p(\mb{x})\, \text{d}\mb{x}
\end{gather}
with $\mb{n}, \mb{Z}$ being calculated from $\mb{\hat{z}}$ and $\mb{m}, \mb{X}$ being calculated from $\mb{x}$.

The question now is how to explicitly calculate $\mb{\hat{z}}$. To do so, we approximate the \ac{gw} distance using a reformulation that would be exact if the covariances were commuting~\cite{Chafai2010}, i.e.,
\begin{align}
\nonumber GW&(\mb{n},\mb{Z}; \mb{m},\mb{X}) \\
\nonumber &= ||\mb{n}-\mb{m}||_2^2 + Tr\{\mb{Z} + \mb{X} - 2(\mb{Z}^{\frac{1}{2}}\mb{X} \mb{Z}^{\frac{1}{2}})^{\frac{1}{2}}\}\\
\nonumber &\approx  ||\mb{n}-\mb{m}||_2^2 + ||\mb{Z}^{\frac{1}{2}} - \mb{X}^{\frac{1}{2}}||_{Frobenius}^2\\
&=||T(\mb{\hat{z}})-T(\mb{x})||_2^2
\end{align}
with transformation
\begin{gather}\label{eqn:transf}
T(\mb{x}) = [m_x\quad m_y\quad s_{(1,1)}\quad s_{(1,2)}\quad  s_{(2,2)}]^{\text{T}}\text{,}\\
\mb{X}^{\frac{1}{2}} = \begin{pmatrix}
s_{(1,1)} & s_{(1,2)}\\
s_{(2,1)} & s_{(2,2)}
\end{pmatrix}\text{.}
\end{gather}
As the square-root of the shape matrix will be symmetric, we include only one corner. Using this reformulation and transformation $T(\cdot)$, we get
\begin{gather}\label{eq_minGW}
\mb{\hat{z}}\approx \underset{\mb{z}}{\mathrm{argmin}}\int||T(\mb{z}) - T(\mb{x})||^2_2\cdot p(\mb{x})\, \text{d}\mb{x}\text{.}
\end{gather}
As the expected value minimizes the mean squared (Euclidean) error, we get the explicit expression 
\begin{gather}\label{eq_minGW}
\mb{\hat{z}}\approx T^{-1}(\mathbb{E}\{T(X)\})
\end{gather}
for the MMGW estimate.
%This is a major insight of this work, as it allows for deriving practical algorithms to calculate the MMGW estimate.

\section{MMGW Fusion}\label{sec_fusion}
In this section, we develop two approaches for the sensor fusion based on MMGW estimation. 
We assume that two (independent) estimates $\mb{\hat{x}}_1$ and $\mb{\hat{x}}_2$  using the parametrization described in Section~\ref{sec_example} are available together with their corresponding covariance matrices $\mb{C}_1$ and $\mb{C}_2$.

In order to fuse these two estimates using the MMGW concept, we have to transform the estimates with the help of $T(\cdot)$ (\ref{eqn:transf}). In the transformed space, the two densities then have to be multiplied before the  MMGW estimate can be calculated  according to  (\ref{eq_minGW}).

Here, we suggest an approximation by using the transformation  $T(\cdot)$ to calculate the means
$\mb{\hat{y}}_1$ and $\mb{\hat{y}}_2$ and  covariance matrices $\mb{P}_1$ and $\mb{P}_2$ in the transformed space.
The fusion  can then be performed in the transformed space based on the Kalman filter formulas, i.e.,
 $T(\mb{\hat{z}})$ becomes
 $$T(\mb{\hat{z}}) =    \mb{P}_2 \cdot (\mb{P}_1 + \mb{P}_2)^{-1} \cdot  \mb{\hat{y}}_1+\mb{P}_1 \cdot (\mb{P}_1 + \mb{P}_2)^{-1}  \cdot \mb{\hat{y}}_2 \enspace . $$
If desired, $\mb{\hat{z}}$ can be explicitly calculated according to
\begin{gather}
\mb{\hat{z}}=T^{-1}(T(\mb{\hat{z}}))\enspace
\end{gather}
with $T^{-1}(\cdot)$ being calculated via the shape matrices eigenvectors and eigenvalues.

%Due to strong nonlinearities, there is no  closed-form solution.

The problem is that the transformation $T(\cdot)$ changes the shape of the state density and introduces a bias. Therefore, we next describe an approach based on linearization in Section~\ref{sec_lin} and a Monte Carlo approach in Section~\ref{sec_particle} to perform the nonlinear transformation $T(\cdot)$. 

\subsection{Linearization}\label{sec_lin}
By utilizing the Jacobians $\mb{H}_i$ for $i\in\{1,2\}$, we get
\begin{align}
\mb{\hat{y}}_i &\approx \mb{H}_i \mb{\hat{x}}_i\enspace,\\
\mb{P}_i&\approx \mb{H}_i\cdot \mb{C}_i \cdot \mb{H}_i^{\text{T}}\enspace.
\end{align}

The calculation of $\mb{H}_i$ can be found in Appendix~\ref{app_jac}. This estimator will be called MMGW-Lin. Unfortunately, the bias introduced by the nonlinear transformation is not dealt with in this linearization. For this reason, a more precise, but also more costly approach is described in the next subsection.

\subsection{Particle Approximation}\label{sec_particle}
To incorporate the bias introduced by the nonlinear transformation $T(\cdot)$, we approximate the transformed density via particles. We will call this estimator MMGW-MC. As the estimates are assumed to be Gaussian with means $\mb{\hat{x}}_i$ and covariances $\mb{C}_i$, we can draw $m$ samples
\begin{gather}
\mb{x}^j_i \sim \mathcal{N}(\mb{\hat{x}}_i, \mb{C}_i) \text{ with } j\in\{1,...,m\}  \text{ and } i\in\{1,2\}\enspace.
\end{gather}

From this, we can use moment approximation to approximate the particle distribution with the first two moments
\begin{align}
\mb{\hat{y}}_i &\approx \frac{1}{m}\underset{j=1}{\overset{m}{\Sigma}}T(\mb{x}_i^j)\enspace,\\
\mb{P}_i &= \frac{1}{m}\underset{j=1}{\overset{m}{\Sigma}}(T(\mb{x}_i^j)-\mb{\hat{y}}_i)\cdot (T(\mb{x}_i^j)-\mb{\hat{y}}_i)^{\text{T}}\enspace.
\end{align}

%Finally, we can fuse the densities in a Kalman fashion and transform the result back into the original state representation.

\section{Alternative: A Heuristic Approach}\label{sec_bestOr}
In this section, we discuss an alternative heuristic approach using the original state representation,
which turned out to perform well with respect to the Gaussian Wasserstein distance.

As mentioned in Section~\ref{sec_example}, the same ellipse can have multiple representations. To be specific, there is an infinite set of parameters which result in the same shape matrix and thus in the same ellipse
\begin{align}
\nonumber T(\alpha, l, w) = T(&\alpha + k\cdot\frac{\pi}{2}, l\cdot p(k) + w\cdot(1-p(k)),\\
&w\cdot p(k) + l\cdot(1-p(k)))\quad\forall k\in\mathbb{Z}\enspace,
\end{align}
with
\begin{gather}
p(k) = \begin{cases}
1 & \quad \text{if } k \text{ is even}\enspace,\\
0 & \quad \text{if } k \text{ is odd}\enspace.\\
\end{cases}
\end{gather}

Using the periodicity of the orientation and replacing $\alpha$ with $\alpha\bmod(2\pi)$, we can reduce it to $4$ possibilities ($k\in\{0, 1, 2, 3\}$). The goal would then be to find the combination for which the fusion holds the highest likelihood. As there should be four equal best combinations, we choose only one version of the first sensor estimate. For example, in the case of two sensors, we choose $\mb{\hat{x}}_1$ and $\mb{\hat{x}}_{2,k}=[m_{x,2}\text{, } m_{y,2}\text{, } \alpha_2 + k\cdot\frac{\pi}{2}\text{, } l_2\cdot p(k) + w_2\cdot(1-p(k))\text{, } w_2\cdot p(k) + l_2\cdot(1-p(k))]^{\text{T}}$ with covariances $\mb{C}_1$ and $\mb{C}_{2,k}$ (rows and columns switched according to $k$), we would get innovation and innovation covariances as
\begin{gather}
\boldsymbol{\nu}_k = \mb{\hat{x}}_1-\mb{\hat{x}}_{2,k}\text{, }
\mb{S}_k = \mb{C}_1+\mb{C}_{2,k}\enspace.
\end{gather}
The goal is then to find the $k_{opt}$ by maximizing the likelihood or minimizing the negative log likelihood according to
\begin{align}
\nonumber k_{opt} = \underset{k}{\mathrm{argmin}}\frac{1}{2}\cdot(&-\boldsymbol{\nu}^{\text{T}}_k \mb{S}_k \boldsymbol{\nu}_k + \log(\det(\mb{S}_k^{-1})) \\
&- 5\log(2\pi))\text{.}
\end{align}
%Then, the estimates can be fused by using $\mb{x}^{(0)}$ and $\mb{x}^{(1)}_{k_{opt}}$.

\section{Evaluation}\label{sec_eval}
\begin{figure*}
\centering
\begin{subfigure}[b]{0.28\textwidth}
\includegraphics[width=\textwidth]{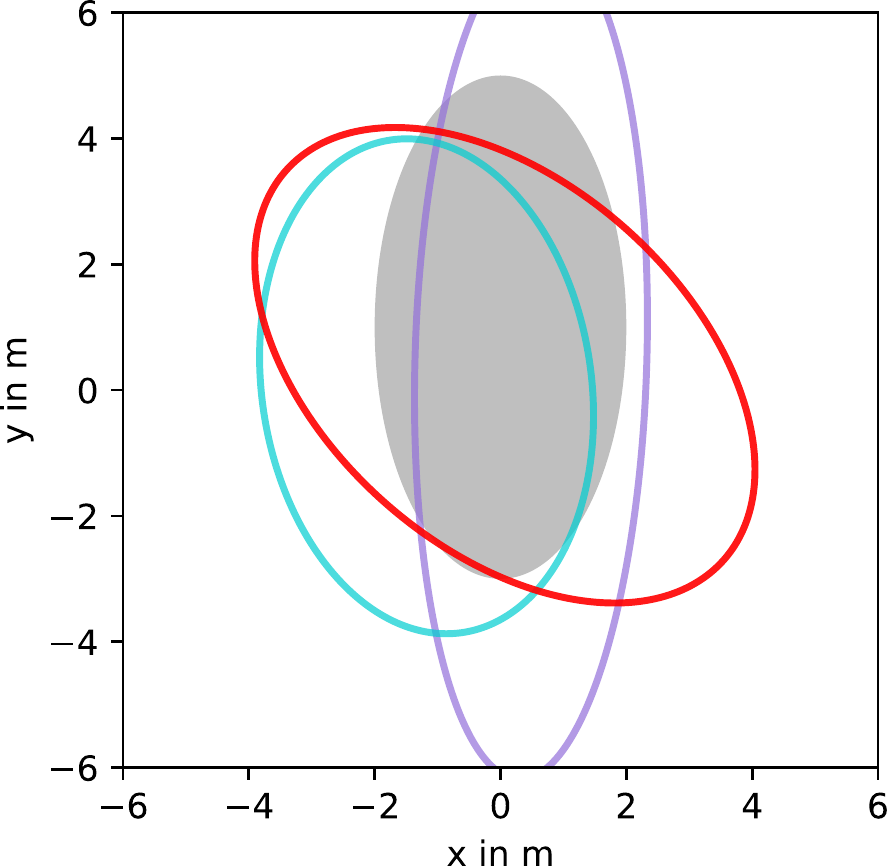}
\caption{Fusion of original state.}
\label{fig_origfus}
\end{subfigure}%
\quad
\begin{subfigure}[b]{0.28\textwidth}
\includegraphics[width=\textwidth]{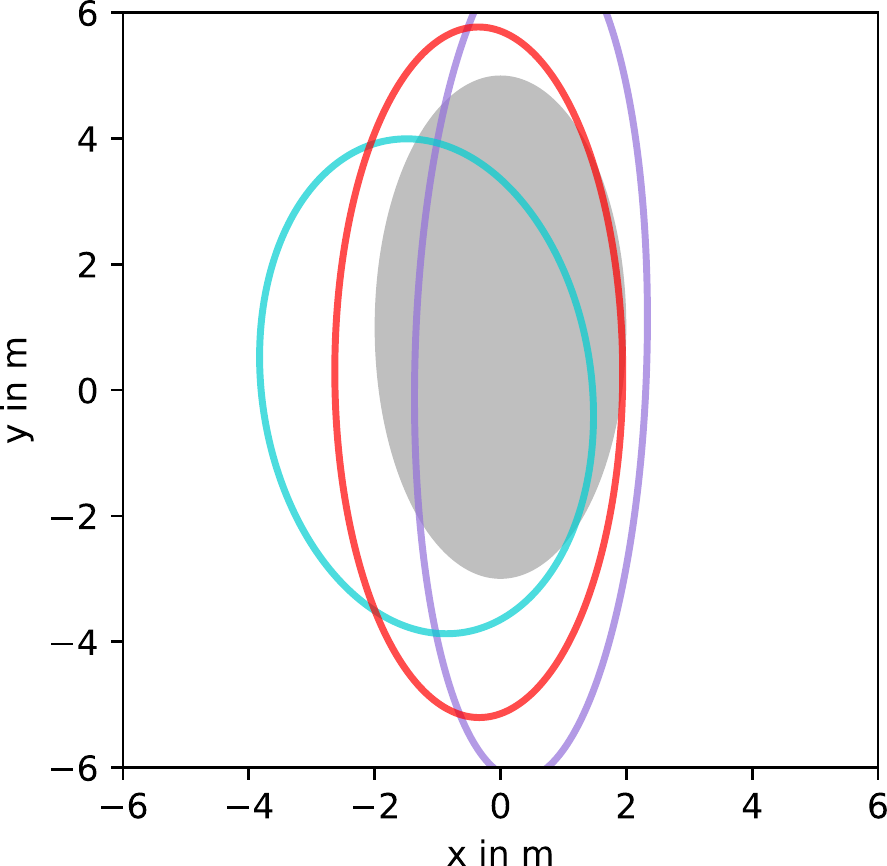}
\caption{Mean of shape matrices.}
\label{fig_shapemean}
\end{subfigure}
\begin{subfigure}[b]{0.28\textwidth}
\includegraphics[width=\textwidth]{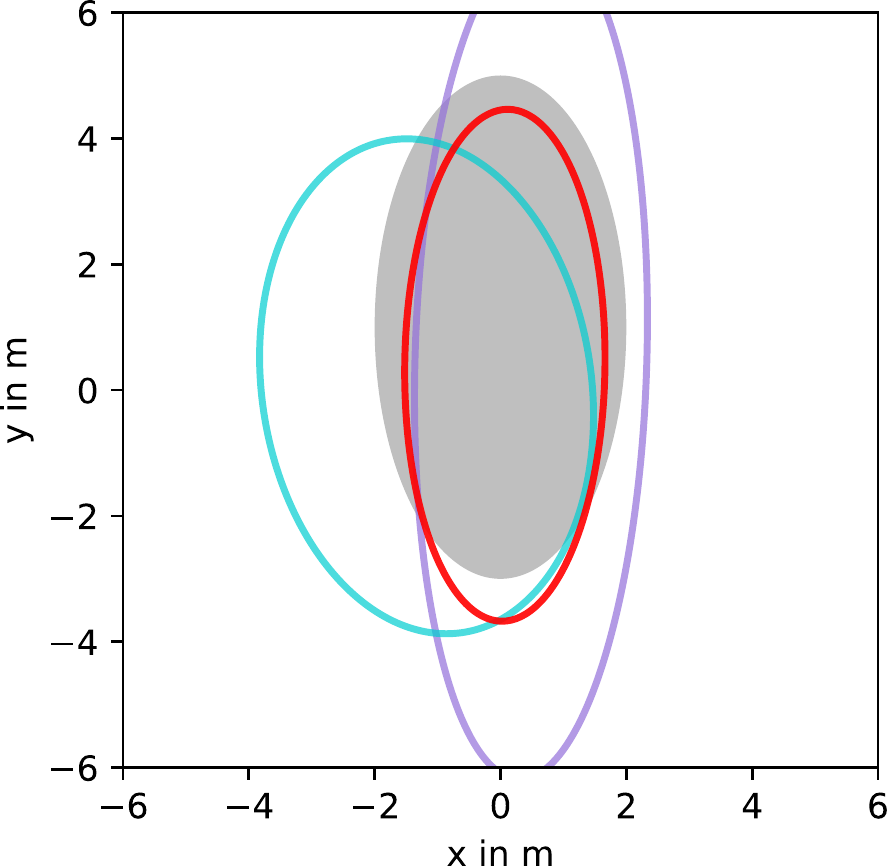}
\caption{MMGW-Lin.}
\label{fig_linfusion}
\end{subfigure}%

\vspace{0.2cm}
\begin{subfigure}[b]{0.28\textwidth}
\includegraphics[width=\textwidth]{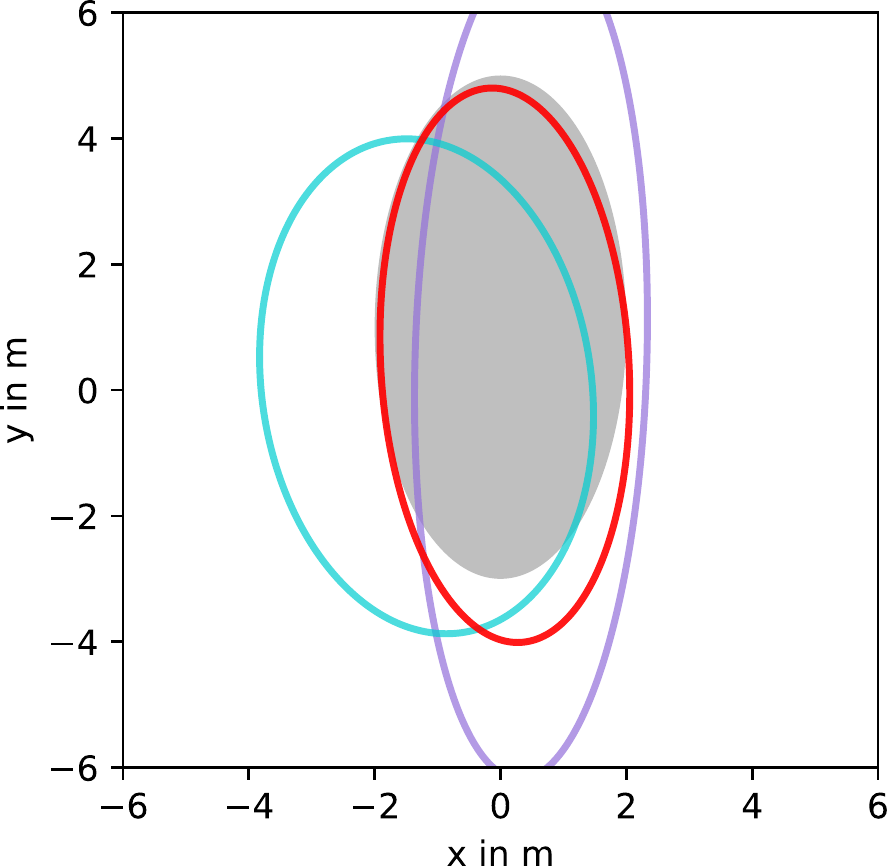}
\caption{Heuristic.}
\label{fig_bestparam}
\end{subfigure}
\begin{subfigure}[b]{0.28\textwidth}
\includegraphics[width=\textwidth]{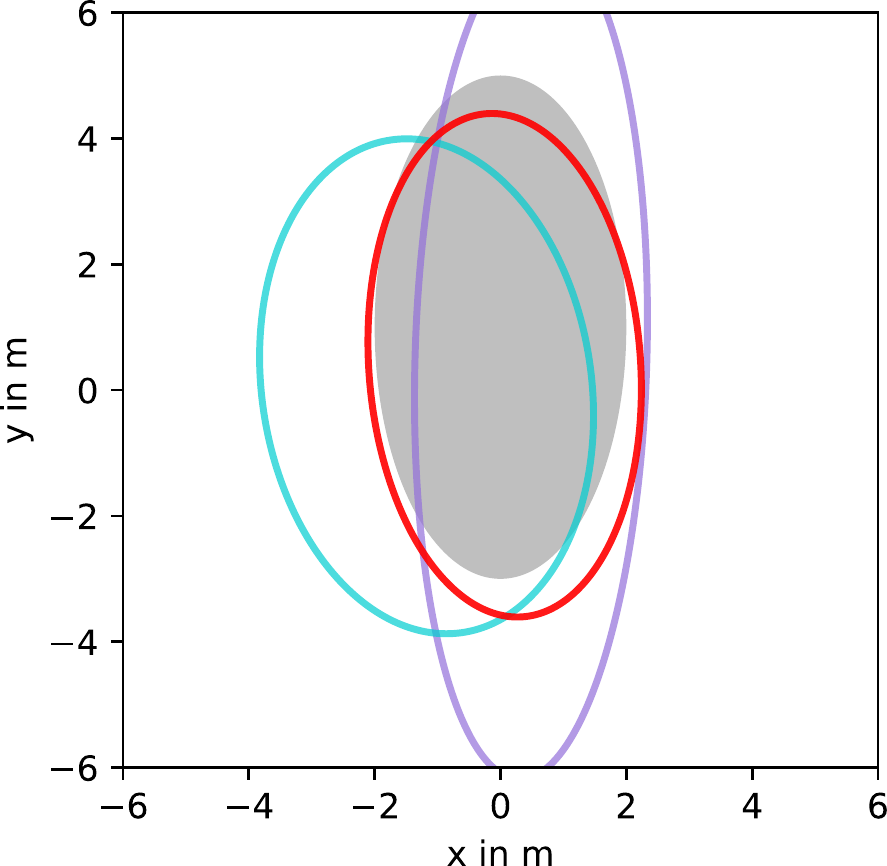}
\caption{MMGW-MC.}
\label{fig_mcfusion}
\end{subfigure}
\caption{Fusion results with ground truth in gray, the two sensor estimates as light blue and purple ellipses and the fusion results in red with the fusion of the original representation in (a), the averaging in (b), the MMGW-Lin in (c), the fusion via the heuristic in (d), and the MMGW-MC in (e).}
\label{fig_fusion}
\end{figure*}

To evaluate our approaches, we simulated a scenario in which two sensors each provide a noisy estimate of an elliptic target. For comparison, we conducted a simple fusion of the original representation of the two estimates (see Figure~\ref{fig_origfus}). Due to not having measurements but getting ellipses directly, we decided to add averaging of the two shape matrices as another alternative to our approach (see Figure~\ref{fig_shapemean}). The results of fusing the transformed state via MMGW-Lin can be seen in Figure~\ref{fig_linfusion} and the results using the heuristic as in Section~\ref{sec_bestOr} are in Figure~\ref{fig_bestparam}. The \mbox{MMGW-MC} with $1000$ samples to approximate the true mean and covariance of the transformed states can be seen in Figure~\ref{fig_mcfusion}.

We evaluated our method in $100$ \ac{mc} runs using two randomly generated ellipses with the ground truth $\mb{x}_g$ as mean and the sensor noises $\mb{C}_1$ and $\mb{C}_2$ as covariances. For the second sensor, we switched $l$ and $w$ and added $\frac{\pi}{2}$ to the orientation (see also the example in Figure~\ref{fig_badexample})
\begin{align}
\mb{x}_g &= [0\text{, }1\text{, }\frac{\pi}{2}\text{, }4\text{, }2]^{\text{T}}\enspace,\\
\mb{C}_1 &= \mathrm{diag}([0.5\text{, }0.5\text{, }0.2\text{, }1\text{, }0.2])^{\text{T}}\enspace,\\
\mb{C}_2 &= \mathrm{diag}([1.5\text{, }1.5\text{, }0.2\text{, }1\text{, }0.2])^{\text{T}}\enspace.
\end{align}

Figure~\ref{fig_fusion} shows an example run. It can be seen that the fusion of the original representation provides the worst results. This is not surprising as the two axes are switched. Averaging the shape matrices is more accurate but still dismisses the known uncertainties for the different parameters. Our fusion methods provide better estimates, especially in scenarios with different noise on the shape parameters. However, due to strong nonlinearities, the linearization using Jacobians does not provide estimates as good as the particle approximation.

\begin{figure}
\centering
\includegraphics[width=0.4\textwidth]{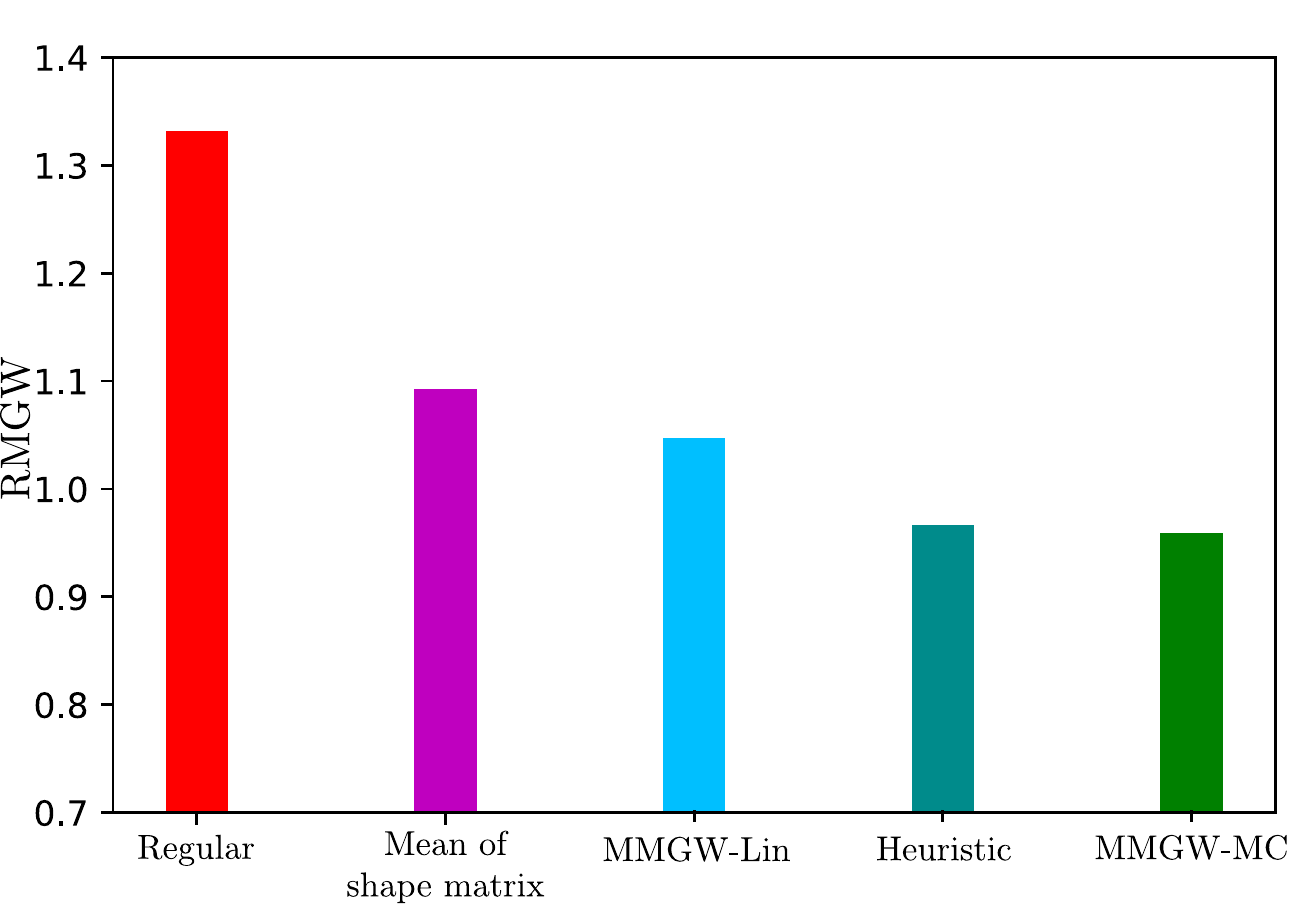}
\caption{Error based on the \ac{gw} distance of (from left to right) the fusion of original parameters (red), mean of shape matrix (magenta), MMGW-Lin (cyan), heuristic (blue), and MMGW-MC (green).}
\label{fig_rmse}
\end{figure}

The results of the $100$ runs can be found in Figure~\ref{fig_rmse} and Table~\ref{tab_rmse}. It can be seen that the naive fusion provides the worst result, followed by the shape mean. The linearization is only slightly better. The heuristic approach provides good results and the MMGW-MC is only a minimal improvement to that. Again, it should be mentioned that the \ac{mc} approximation involves a Gaussian approximation, meaning even better results would be obtained by means of a direct fusion of the densities.

\begin{table}[h]
    \centering
    \begin{tabular}{c|c}
    Method      & RMGW\\
    \hline
    Regular     & 1.3316\\
    Mean of shape matrix  & 1.0924\\
    MMGW-Lin   & 1.0470\\
    Heuristic & 0.9661\\
    MMGW-MC       & 0.9590
    \end{tabular}
    \caption{Error based on the \ac{gw} distance of the approaches.}
    \label{tab_rmse}
\end{table}

\section{Conclusion and Future Work}\label{sec_concl}
In this work, we have proposed a systematic approach to fuse extended target estimates.
The central concept is to derive estimators that are optimal with respect to a metric on shapes -- here the Gaussian Wasserstein distance.
% A major insight is that the Minimum Mean Gaussian Wasserstein  (MMGW) estimate can be derived explicitly.
This was achieved  by an (approximate) reformulation of the \ac{gw} distance to find a transformation of the original state space into a space in which the minimization of the Euclidean distance   minimizes (an approximation of the) \ac{gw} distance. 

Based on the MMGW estimation philosophy, we derived practical fusion methods for elliptic extended target estimates. We applied Monte Carlo and linearization techniques to approximate the moments of the transformed estimates. By this means, the estimates could be fused in Kalman fashion. We also proposed a heuristic approach to test a minimum number of representations of the same ellipse to fuse the estimates in the original state space.

For future work, we aim at considering the fusion of other shapes, like rectangles or more arbitrary ones~\cite{Fusion11_Baum-RHM}. Furthermore, we  seek to evaluate
 the approximation quality of the proposed estimator with respect to a direct optimization of the mean GW distance.

\appendix
\subsection{Appendix I}\label{app_jac}
To calculate Jacobian $\mb{H}$, define the transformation as $T(\cdot)=h(\mb{x})=[m_x\text{, }m_y\text{, }h_3(\mb{x})\text{, }h_4(\mb{x})\text{, }h_5(\mb{x})]^{\text{T}}$ (the sensor notation is omitted for readability) %TODO add citations?
\begin{align}
h_3(\mb{x}) &= t^{-1}(t_{(11)}+s)\enspace,\\
h_4(\mb{x}) &= t^{-1}t_{(1,2)}\enspace,\\
h_5(\mb{x}) &= t^{-1}(t_{(2,2)} + s)\enspace,\\
t &= \sqrt{t_{(1,1)}+t_{(2,2)}+s}\enspace,\\
s &= 2\sqrt{t_{(1,1)}t_{(2,2)}-t_{(1,2)}t_{(2,1)}}\enspace,
\end{align}
with $t_{(k,l)}$ being the elements of the shape matrix
\begin{align}
t_{(1,1)} &= l^2\cos(\alpha)^2 + w^2\sin(\alpha)^2\enspace,\\
t_{(1,2)} &= (l^2-w^2)\cos(\alpha)\sin(\alpha)\enspace,\\
t_{(2,1)} &= t_{(1,2)}\enspace,\\
t_{(2,2)} &= l^2\sin(\alpha)^2+w^2\cos(\alpha)^2\enspace.
\end{align}
\begin{align}
\mb{H} &= \begin{pmatrix}
1 & 0 & 0 & 0 & 0\\
0 & 1 & 0 & 0 & 0\\
0 & 0 & \frac{\partial h_3}{\partial \alpha} & \frac{\partial h_3}{\partial l} & \frac{\partial h_3}{\partial w}\\
0 & 0 & \frac{\partial h_4}{\partial \alpha} & \frac{\partial h_4}{\partial l} & \frac{\partial h_4}{\partial w}\\
0 & 0 & \frac{\partial h_5}{\partial \alpha} & \frac{\partial h_5}{\partial l} & \frac{\partial h_5}{\partial w}
\end{pmatrix} \enspace ,
\end{align}
\begin{align}
\frac{\partial h_3}{\partial u} &= -t^{-2}\frac{\partial t}{\partial u}(t_{(1,1)}+s)+t^{-1}(\frac{\partial t_{(1,1)}}{\partial u}+\frac{\partial s}{\partial u})\\
\frac{\partial h_4}{\partial u} &= -t^{-2}\frac{\partial t}{\partial u}t_{(1,2)}+t^{-1}\frac{\partial t_{(1,2)}}{\partial u}\\
\frac{\partial h_5}{\partial u} &= -t^{-2}\frac{\partial t}{\partial u}(t_{(2,2)}+s)+t^{-1}(\frac{\partial t_{(2,2)}}{\partial u}+\frac{\partial s}{\partial u})\\
\frac{\partial t}{\partial u} &= \frac{\frac{\partial t_{(1,1)}}{\partial u}+\frac{\partial t_{(2,2)}}{\partial u}+\frac{\partial s}{\partial u}}{2t}\\
\frac{\partial s}{\partial u} &= \frac{\frac{\partial t_{(1,1)}}{\partial u}t_{(2,2)}+t_{(1,1)}\frac{\partial t_{(2,2)}}{\partial u}-2\frac{\partial t_{(1,2)}}{\partial u}t_{(1,2)}}{\sqrt{t_{(1,1)}t_{(2,2)}-t_{(1,2)}^2}}\\
\frac{\partial t_{(1,1)}}{\partial \alpha} &= (w^2-l^2)\sin(2\alpha)\\
\frac{\partial t_{(1,1)}}{\partial l} &= 2l\cos(\alpha)^2\\
\frac{\partial t_{(1,1)}}{\partial w} &= 2w\sin(\alpha)^2
\end{align}
\begin{align}
\nonumber\\
\frac{\partial t_{(1,2)}}{\partial \alpha} &= (l^2-w^2)\cos(2\alpha)\\
\frac{\partial t_{(1,2)}}{\partial l} &= 2l\cos(\alpha)\sin(\alpha)\\
\frac{\partial t_{(1,2)}}{\partial w} &= -2w\sin(\alpha)\cos(\alpha)\\
\frac{\partial t_{(2,2)}}{\partial \alpha} &= (l^2-w^2)\sin(2\alpha)\\
\frac{\partial t_{(2,2)}}{\partial l} &= 2l\sin(\alpha)^2\\
\frac{\partial t_{(2,2)}}{\partial w} &= 2w\cos(\alpha)^2
\end{align}


%\ifCLASSOPTIONcompsoc
%  % The Computer Society usually uses the plural form
%\else
%  % regular IEEE prefers the singular form
%\fi



%\bibliographystyle{IEEEtran}

\bibliographystyle{IEEEtran}
%\bibliography{IEEEabrv,lit}
\bibliography{./Literature,./publicationsFusion,./thesesFusion}

% Generated by IEEEtranS.bst, version: 1.12 (2007/01/11)
\begin{thebibliography}{10}
\providecommand{\url}[1]{#1}
\csname url@samestyle\endcsname
\providecommand{\newblock}{\relax}
\providecommand{\bibinfo}[2]{#2}
\providecommand{\BIBentrySTDinterwordspacing}{\spaceskip=0pt\relax}
\providecommand{\BIBentryALTinterwordstretchfactor}{4}
\providecommand{\BIBentryALTinterwordspacing}{\spaceskip=\fontdimen2\font plus
\BIBentryALTinterwordstretchfactor\fontdimen3\font minus
  \fontdimen4\font\relax}
\providecommand{\BIBforeignlanguage}[2]{{%
\expandafter\ifx\csname l@#1\endcsname\relax
\typeout{** WARNING: IEEEtranS.bst: No hyphenation pattern has been}%
\typeout{** loaded for the language `#1'. Using the pattern for}%
\typeout{** the default language instead.}%
\else
\language=\csname l@#1\endcsname
\fi
#2}}
\providecommand{\BIBdecl}{\relax}
\BIBdecl

\bibitem{Agueh2011}
M.~Agueh and G.~Carlier, ``{Barycenters in the Wasserstein Space},'' \emph{SIAM
  Journal on Mathematical Analysis}, vol.~43, no.~2, pp. 904--924, 2011.

\bibitem{Fusion11_Baum-RHM}
M.~Baum and U.~D. Hanebeck, ``{Shape Tracking of Extended Objects and Group
  Targets with Star-Convex RHMs},'' in \emph{Proceedings of the 14th
  International Conference on Information Fusion (Fusion 2011)}, Chicago,
  Illinois, USA, Jul. 2011.

\bibitem{J_TAES_Baum_RHM}
------, ``{Extended Object Tracking with Random Hypersurface Models},''
  \emph{IEEE Transactions on Aerospace and Electronic Systems}, vol.~50, pp.
  149--159, Jan. 2014.

\bibitem{J_SPL15_Baum}
M.~Baum, P.~Willett, and U.~D. Hanebeck, ``{On Wasserstein Barycenters and
  MMOSPA Estimation},'' \emph{IEEE Signal Processing Letters}, vol.~22, no.~10,
  pp. 1511--1515, Oct. 2015.

\bibitem{Bigot2012}
J.~{Bigot} and T.~{Klein}, ``{Consistent Estimation of a Population Barycenter
  in the Wasserstein Space},'' \emph{ArXiv e-prints}, Dec. 2012.

\bibitem{Brosseit2017}
P.~Brosseit, B.~Duraisamy, and J.~Dickmann, ``{The Volcanormal Density for
  Radar-based Extended Target Tracking},'' in \emph{2017 IEEE 20th
  International Conference on Intelligent Transportation Systems (ITSC)}, Oct
  2017, pp. 1--6.

\bibitem{Carlier2014}
G.~Carlier, A.~Oberman, and E.~Oudet, ``{Numerical Methods for Matching for
  Teams and Wasserstein Barycenters},'' Tech. Rep., May 2014.

\bibitem{Chafai2010}
\BIBentryALTinterwordspacing
D.~Chafa\"i, ``{Wasserstein Distance Between Two Gaussians},'' April 30 2010,
  {Accessed: 1-6-2019}. [Online]. Available:
  \url{http://djalil.chafai.net/blog/2010/04/30/wasserstein-distance-between-two-gaussians/}
\BIBentrySTDinterwordspacing

\bibitem{Crouse2011e}
D.~F. Crouse, P.~Willett, Y.~Bar-Shalom, and L.~Svensson, ``{Aspects of MMOSPA
  Estimation},'' in \emph{Proceedings of the 50th IEEE Conference on Decision
  and Control and European Control Conference, Orlando, FL}, Dec. 2011.

\bibitem{Crouse2013}
D.~F. Crouse, ``{Advances in Displaying Uncertain Estimates of Multiple
  Targets},'' in \emph{SPIE -- Signal Processing, Sensor Fusion, and Target
  Recognition XXII}, vol. 8745, 2013, pp. 874\,504--874\,504--31.

\bibitem{Cuturi2013}
M.~{Cuturi} and A.~{Doucet}, ``{Fast Computation of Wasserstein Barycenters},''
  in \emph{Proceedings of the 31st International Conference on Machine Learning
  (ICML-14)}, Oct. 2014.

\bibitem{Duraisamy2016track}
B.~Duraisamy, M.~Gabb, A.~V. Nair, T.~Schwarz, and T.~Yuan, ``{Track Level
  Fusion of Extended Objects from Heterogeneous Sensors},'' in
  \emph{Proceedings of the 19th International Conference on Information Fusion
  (Fusion 2016)}.\hskip 1em plus 0.5em minus 0.4em\relax IEEE, 2016, pp.
  876--885.

\bibitem{Durrant2008multisensor}
H.~Durrant-Whyte and T.~C. Henderson, ``{Multi Sensor Data Fusion},''
  \emph{Springer handbook of robotics}, pp. 585--610, 2008.

\bibitem{Feldmann2010}
M.~Feldmann, D.~Fr{\"{a}}nken, and W.~Koch, ``{Tracking of Extended Objects and
  Group Targets using Random Matrices},'' \emph{IEEE Transactions on Signal
  Processing}, vol.~59, no.~4, pp. 1409--1420, 2011.

\bibitem{Frohle2019decentralized}
M.~Fr{\"o}hle, K.~Granstr{\"o}m, and H.~Wymeersch, ``{Decentralized Poisson
  Multi-Bernoulli Filtering for Extended Target Tracking},'' \emph{arXiv
  preprint arXiv:1901.04518}, 2019.

\bibitem{Givens1984}
C.~R. Givens and R.~M. Shortt, ``{A Class of Wasserstein Metrics for
  Probability Distributions},'' \emph{The Michigan Mathematical Journal},
  vol.~31, no.~2, pp. 231--240, 1984.

\bibitem{Granstroem2017}
K.~Granstr\"om, M.~Baum, and S.~Reuter, ``{Extended Object Tracking:
  Introduction, Overview and Applications},'' \emph{ISIF Journal of Advances in
  Information Fusion}, vol.~12, no.~2, Dec. 2017.

\bibitem{Granstroem2011}
K.~Granstr{\"{o}}m, C.~Lundquist, and U.~Orguner, ``{Tracking Rectangular and
  Elliptical Extended Targets Using Laser Measurements},'' in \emph{Proceedings
  of the 14th International Conference on Information Fusion (Fusion 2011)},
  Chicago, Illinois, USA, Jul. 2011.

\bibitem{granstrom2013spawning}
K.~Granstr\"om and U.~Orguner, ``{On Spawning and Combination of Extended/Group
  Targets Modeled With Random Matrices},'' \emph{IEEE Transactions on Signal
  Processing}, vol.~61, no.~3, pp. 678--692, 2013.

\bibitem{Guerriero2010}
M.~Guerriero, L.~Svensson, D.~Svensson, and P.~Willett, ``{Shooting Two Birds
  with Two Bullets: How to Find Minimum Mean OSPA Estimates},''
  \emph{Proceedings of the 13th International Conference on Information Fusion
  (Fusion 2010)}, 2010.

\bibitem{Hirscher2016}
T.~Hirscher, A.~Scheel, S.~Reuter, and K.~Dietmayer, ``{Multiple Extended
  Object Tracking using Gaussian Processes},'' in \emph{Proceedings of the 19th
  International Conference on Information Fusion (Fusion 2016)}, Jul. 2016, pp.
  868--875.

\bibitem{Kaulbersch2018_Fusion}
H.~Kaulbersch, J.~Honer, and M.~Baum, ``{A Cartesian {B-Spline} Vehicle Model
  for Extended Object Tracking},'' in \emph{21st International Conference on
  Information Fusion (FUSION 2018)}, Jul. 2018.

\bibitem{Koch2008}
W.~Koch, ``{Bayesian Approach to Extended Object and Cluster Tracking using
  Random Matrices},'' \emph{IEEE Transactions on Aerospace and Electronic
  Systems}, vol.~44, no.~3, pp. 1042--1059, Jul. 2008.

\bibitem{Lan2016b}
J.~Lan and X.~R. Li, ``{Tracking of Extended Object or Target Group Using
  Random Matrix: New Model and Approach},'' \emph{IEEE Transactions on
  Aerospace and Electrical Systems}, vol.~52, no.~6, pp. 2973--2988, Dec. 2016.

\bibitem{Michaelis2017heterogeneous}
M.~Michaelis, P.~Berthold, D.~Meissner, and H.-J. Wuensche, ``{Heterogeneous
  Multi-Sensor Fusion for Extended Objects in Automotive Scenarios Using
  Gaussian Processes and a GMPHD-Filter},'' in \emph{2017 Sensor Data Fusion:
  Trends, Solutions, Applications (SDF)}.\hskip 1em plus 0.5em minus
  0.4em\relax IEEE, 2017, pp. 1--6.

\bibitem{Mihaylova2014}
L.~Mihaylova, A.~Carmi, F.~Septier, A.~Gning, S.~Pang, and S.~Godsill,
  ``{Overview of Bayesian Sequential Monte Carlo Methods for Group and Extended
  Object Tracking},'' \emph{Digital Signal Processing}, vol.~25, pp. 1--16,
  Feb. 2014.

\bibitem{Nilsson2016}
S.~Nilsson and A.~Klekamp, ``{Object Level Fusion of Extended Dynamic
  Objects},'' in \emph{2016 IEEE International Conference on Multisensor Fusion
  and Integration for Intelligent Systems (MFI)}, Sept 2016, pp. 251--258.

\bibitem{Rabin2012}
J.~Rabin, G.~Peyre, J.~Delon, and M.~Bernot, ``{Wasserstein Barycenter and Its
  Application to Texture Mixing},'' in \emph{Scale Space and Variational
  Methods in Computer Vision}, ser. Lecture Notes in Computer Science.\hskip
  1em plus 0.5em minus 0.4em\relax Springer Berlin Heidelberg, 2012, vol. 6667,
  pp. 435--446.

\bibitem{Schuhmacher2008}
D.~Schuhmacher, B.-T. Vo, and B.-N. Vo, ``{A Consistent Metric for Performance
  Evaluation of Multi-Object Filters},'' \emph{IEEE Transactions on Signal
  Processing}, vol.~56, no.~8, pp. 3447 --3457, Aug. 2008.

\bibitem{Teich2017}
F.~Teich, S.~Yang, and M.~Baum, ``{GM-PHD filter for Multiple Extended Object
  Tracking based on the Multiplicative Error Shape Model and Network Flow
  Labeling},'' in \emph{Proceedings of the IEEE Intelligent Vehicles Symposium
  (IV 2017)}, Redondo Beach, CA, USA, Jun. 2017.

\bibitem{Veltkamp2000}
R.~Veltkamp and M.~Hagedoorn, ``Shape similarity measures, properties and
  constructions,'' in \emph{Advances in Visual Information Systems}, ser.
  Lecture Notes in Computer Science, R.~Laurini, Ed.\hskip 1em plus 0.5em minus
  0.4em\relax Springer Berlin / Heidelberg, 2000, vol. 1929, pp. 133--153.

\bibitem{Vivone2015}
G.~Vivone, P.~Braca, K.~Gran\-str{\"{o}}m, A.~Natale, and J.~Chanussot,
  ``{Converted Measurements Random Matrix Approach to Extended Target Tracking
  Using X-band Marine Radar Data},'' in \emph{18th International Conference on
  Information Fusion (Fusion 2015)}, Washington, DC, USA, Jul. 2015, pp.
  976--983.

\bibitem{Vivone2017}
G.~Vivone, K.~Granstr{\"o}m, P.~Braca, and P.~Willett, ``{Multiple Sensor
  Measurement Updates for the Extended Target Tracking Random Matrix Model},''
  \emph{IEEE Transactions on Aerospace and Electronic Systems}, vol.~53, no.~5,
  pp. 2544--2558, 2017.

\bibitem{Vo2015}
B.-N. Vo, M.~Mallick, Y.~Bar-Shalom, S.~Coraluppi, R.~Osborne, R.~Mahler, B.-T.
  Vo, and J.~G. Webster, \emph{{Multitarget Tracking}}.\hskip 1em plus 0.5em
  minus 0.4em\relax John Wiley \& Sons, Inc., 2015.

\bibitem{wahlstrom2015extended}
N.~Wahlstr{\"o}m and E.~{\"O}zkan, ``{Extended Target Tracking Using Gaussian
  Processes},'' \emph{IEEE Transactions on Signal Processing}, vol.~63, no.~16,
  pp. 4165--4178, 2015.

\bibitem{Yang2018_arXiv}
S.~Yang and M.~Baum, ``{Tracking the Orientation and Axes Lengths of an
  Elliptical Extended Object},'' \emph{CoRR}, vol. abs/1805.03276v1, May 2018.

\bibitem{MFI16_Yang}
S.~Yang, M.~Baum, and K.~Granstr\"om, ``{Metrics for Performance Evaluation of
  Elliptic Extended Object Tracking Methods},'' in \emph{Proceedings of the
  2016 IEEE International Conference on Multisensor Fusion and Integration for
  Intelligent Systems (MFI 2016)}, Baden-Baden, Germany, Sep. 2016.

\end{thebibliography}

\end{document}